\documentclass{article}
\usepackage{PRIMEarxiv}
\usepackage[utf8]{inputenc} 
\usepackage[T1]{fontenc}    
\usepackage{hyperref}       
\usepackage{url}            
\usepackage{booktabs}       
\usepackage{amsfonts}       
\usepackage{nicefrac}       
\usepackage{microtype}      
\usepackage{lipsum}
\usepackage{fancyhdr}       
\usepackage{graphicx}       
\graphicspath{{figs/}}      
\usepackage{float}
\usepackage{hyperref}
\usepackage{tcolorbox}
\usepackage{enumitem}

\usepackage[style=apa]{biblatex}[2023/03/05]
\title{
Generative AI Enhanced Financial Risk Management Information Retrieval\\
}

\author{
  Amin Haeri, Jonathan Vitrano, Mahdi Ghelichi\\
  Model Development Innovation, Risk Management\\
  TD Bank, Toronto, Canada\\
  \texttt{\{amin.haeri, jonathan.vitrano, mahdi.ghelichi\}@td.com}
}

\begin{document}
\maketitle

\begin{abstract}
Risk management in finance involves recognizing, evaluating, and addressing financial risks to maintain stability and ensure regulatory compliance. Extracting relevant insights from extensive regulatory documents is a complex challenge requiring advanced retrieval and language models. This paper introduces RiskData\footnote{\href{https://huggingface.co/datasets/aminhaeri/risk-data}{https://huggingface.co/datasets/aminhaeri/risk-data}}, a dataset specifically curated for finetuning embedding models in risk management, and RiskEmbed\footnote{\href{https://huggingface.co/aminhaeri/risk-embed}{https://huggingface.co/aminhaeri/risk-embed}}, a finetuned embedding model designed to improve retrieval accuracy in financial question-answering systems. The dataset is derived from 94 regulatory guidelines published by the Office of the Superintendent of Financial Institutions (OSFI) from 1991 to 2024. We finetune a state-of-the-art sentence BERT embedding model to enhance domain-specific retrieval performance typically for Retrieval-Augmented Generation (RAG) systems. Experimental results demonstrate that RiskEmbed significantly outperforms general-purpose and financial embedding models, achieving substantial improvements in ranking metrics. By open-sourcing both the dataset and the model, we provide a valuable resource for financial institutions and researchers aiming to develop more accurate and efficient risk management AI solutions.
\end{abstract}


\section{Introduction} \label{sec:intro}

Financial risk management is a cornerstone of maintaining a robust and stable financial system. It is broadly defined as the process of identifying, assessing, and mitigating potential financial risks that could jeopardize an organization's assets and operational stability \parencite{risk1}. This proactive approach is crucial for financial institutions as it enables them to navigate the inherent uncertainties of financial markets and protect themselves against adverse events \parencite{risk2}. Effective risk management encompasses a spectrum of potential threats, broadly categorized into Market Risk, which arises from fluctuations in market prices; Credit Risk, stemming from the possibility of borrower default; and Liquidity Risk, which refers to the inability to meet short-term financial obligations when due \parencite{risk3}. By diligently managing these interconnected risks, financial institutions aim to ensure their solvency, maintain stakeholder confidence, and ultimately contribute to the overall health and resilience of the financial ecosystem.

To effectively manage these diverse risks, financial institutions operate within structured regulatory frameworks, most notably those established by international bodies such as the Basel Committee on Banking Supervision (BCBS). The Basel Accords, particularly Basel III, provide comprehensive guidelines designed to strengthen banking regulation, supervision, and risk management practices globally \parencite{BCBS2019}. These frameworks mandate specific capital adequacy ratios, liquidity coverage ratios, and stress testing protocols, among other requirements, to enhance the resilience of financial institutions to economic shocks and systemic crises \parencite{BIS2017}. However, the sheer volume and complexity of these regulatory documents, coupled with their frequent updates and interpretations, present a significant challenge for institutions seeking to ensure full compliance and extract actionable insights. This complexity necessitates the adoption of advanced analytical tools and the development of specialized expertise to efficiently process, interpret, and implement these intricate regulatory mandates, highlighting the ongoing need for innovation in risk management methodologies and technologies.

Natural Language Processing (NLP) has become a powerful tool for analyzing domain-specific data at scale. The financial domain uses NLP to translate varied data for rapid decision-making. For example, sentiment analysis in finance has proven to be effective in predicting stock market behavior and credit risks \parencite{NLPApp1, NLPApp2, NLPApp3, NLPApp4}. This encourages additional investigation into utilizing NLP techniques such as Question Answering (QA) within the field of risk management (as the central theme of this paper). These systems can assist experts in quickly retrieving the specific information they need, improving efficiency and decision-making and yielding considerable profits.

Large Language Models (LLMs) are advanced artificial intelligence systems based on the Transformer architecture, which enables them to process and generate text by leveraging self-attention mechanisms and deep neural networks \parencite{transformer}. While LLMs are not inherently QA systems, they can be used as one. At their core, LLMs are sequence-to-sequence models trained on text data to predict the most likely next token in a sequence. This allows them to generate text, complete sentences, summarize content, and even engage in conversational dialogue. They have revolutionized NLP by leveraging massive datasets and billions of parameters to generate human-like text, answer queries, and perform complex language-related tasks. However, despite their impressive capabilities, LLMs face critical challenges. One major issue is hallucination, where the model generates plausible-sounding but incorrect or misleading information \parencite{LLMHall}. Additionally, LLMs struggle with domain-specific knowledge, as they are trained on broad datasets that may not include up-to-date or specialized content required for particular industries like finance, healthcare, or law \parencite{LLMLimit1}. Another key limitation is their static nature—once trained, they cannot dynamically incorporate new knowledge without expensive retraining \parencite{LLMLimit2}.

\begin{figure}[!tb]
  \centering
  \includegraphics[width=1\textwidth]{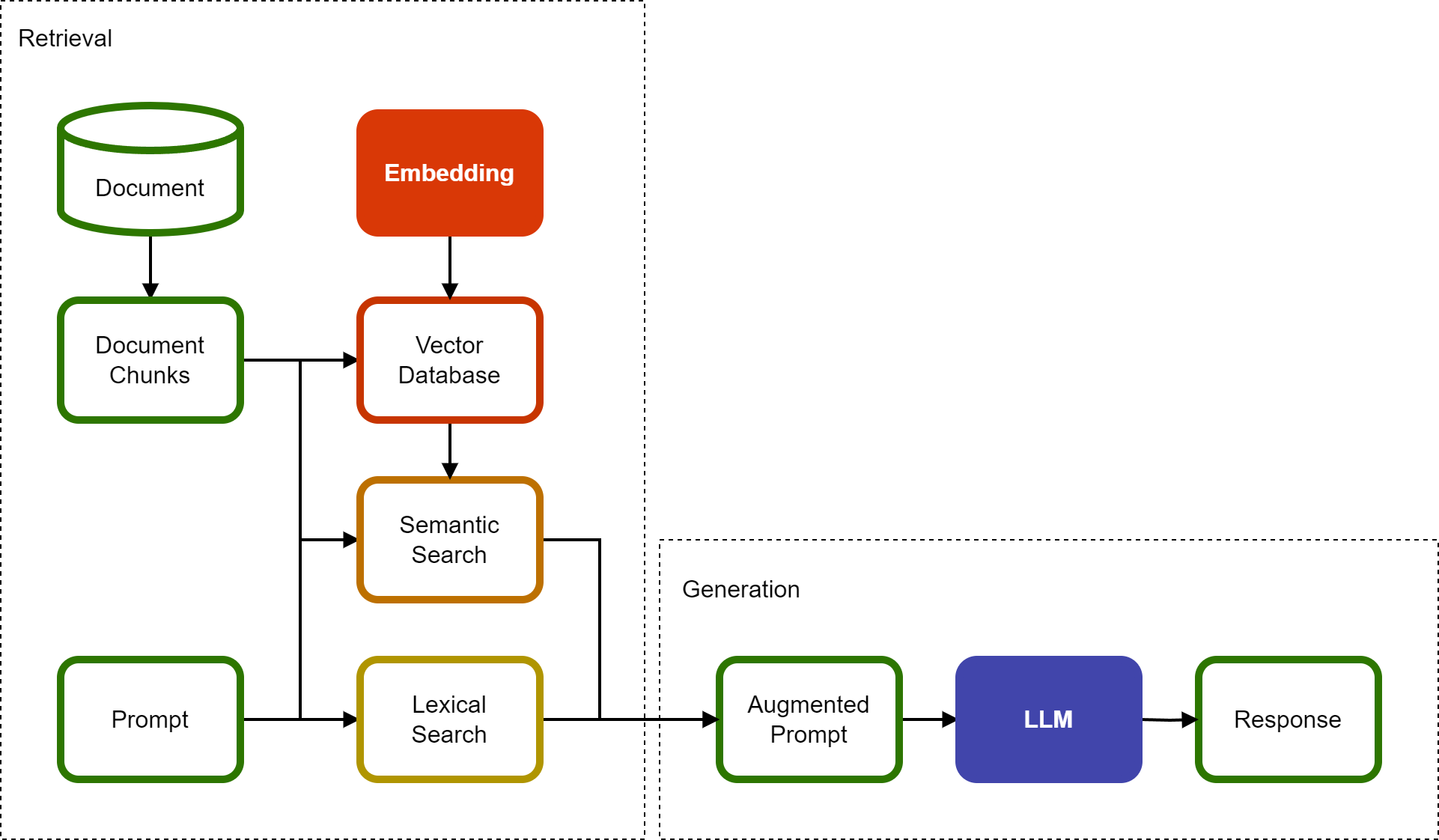}
  \caption{
  Overview of the Retrieval-Augmented Generation (RAG) pipeline consisting of multiple information retrieval components and a generative large language model. The embedding model is crucial in transforming document chunks into dense vector embeddings, enabling efficient semantic search and retrieval of relevant information for prompt augmentation.}
  \label{fig:rag}
\end{figure}
To address these issues, Retrieval-Augmented Generation (RAG) has emerged as an effective and efficient solution for domain-specific applications \parencite{Rag}. RAG enhances LLMs by integrating an information retrieval (IR) system that fetches relevant documents or structured data at query time, enabling the model to generate more factually accurate and contextually relevant responses \parencite{RagExp}. This approach reduces hallucinations by grounding responses in authoritative sources and allows LLMs to stay updated without full retraining \parencite{RagHall1, RagHall2, RagInfo}. By combining retrieval with generation, RAG not only improves response accuracy but also offers better interpretability and control, making it a practical and scalable solution for real-world enterprise applications \parencite{RagScale}. A typical RAG system consists of multiple IR components that work with a generative model (i.e., LLM) shown in Figure \ref{fig:rag}. A document parser processes unstructured text, making it retrievable. An embedding model converts text into vector representations, enabling semantic search using a vector database. To improve retrieval, hybrid search combines lexical search for exact matches with semantic search for contextual understanding. Note that retrieved documents can further be refined through re-ranking, where transformer-based models prioritize the most relevant results \parencite{rerank1, rerank2}. Finally, the LLM generates a response using the prompt and retrieved documents (i.e., augmented prompt), reducing hallucinations and improving domain-specific accuracy.

Transformer-based embedding models, such as BERT (Bidirectional Encoder Representations from Transformers) \parencite{bert} and sentence BERT (sBERT) \parencite{sbert}, are the core of modern information retrieval systems. These models are designed to understand the context of words within a sentence by processing the input text bidirectionally, meaning they consider both the preceding and following words to capture deeper semantic meaning. This bidirectional approach allows transformers to go beyond simple word associations and capture more nuanced aspects of language, such as word sense disambiguation (WSD) \parencite{WDS, transformer}. When used for embedding, a transformer model generates dense vector representations of text (words, sentences, or documents) that capture not only syntactic meaning but also the intent, context, and relationships between words \parencite{WDR}. While BERT processes individual tokens and learns contextual embeddings for each word in a sentence, sBERT is specifically optimized to generate embeddings for entire sentences or paragraphs. sBERT, which builds upon BERT's architecture, is finetuned to produce fixed-size dense vectors that represent the overall meaning of a sentence or document, making them ideal for tasks such as semantic similarity, clustering, and information retrieval \parencite{sbert, distilbert}. Unlike BERT, which focuses on token-level understanding, sBERT uses techniques such as mean or max pooling to condense token-level information into a single, unified representation for a sentence. This makes sBERT particularly effective in scenarios that require comparing or retrieving entire pieces of text based on their semantic content.

In RAG systems, as discussed above, sBERT embedding models allow for effective semantic search, where queries and documents are converted into vector representations that can be compared using similarity measures. The embedding model enables the system to understand the underlying meaning of both the query and the document, making it the core component for accurate and efficient retrieval. However, a pre-trained embedding model is generally trained on broad and diverse text sources and may not capture the specialized terminology, context, or nuances specific to a particular domain, such as risk management \parencite{roberta, biobert}. This limitation can result in suboptimal performance when the model is applied to domain-specific tasks, as it may not accurately understand or retrieve the most relevant documents that involve complex financial terminology, regulatory terms, or risk-related concepts. For instance, phrases commonly used in risk management, such as "credit exposure" and "stress testing" might not be well represented in a generic pre-trained model's vocabulary, leading to inaccurate retrieval or poor semantic matching. To address this limitation, and given the inherent parameter constraints of embedding models often imposed by latency requirements, finetuning these models on a domain-specific dataset becomes a critical step. In the case of risk management, this would involve training the model on risk-related texts, such as financial reports, regulatory documents, or industry publications, so that the embeddings better reflect the specific vocabulary and context of the domain. Finetuning the model ensures that the information retrieval system accurately captures the specialized knowledge needed to improve the overall performance of RAG systems in risk management applications.



This work addresses the critical need for domain-specific embedding models in risk management by introducing a novel QA dataset tailored for targeted finetuning. In addition, it presents a finetuned embedding model, open-sourced for community use, and demonstrates its superior performance on risk management tasks through rigorous benchmarking against state-of-the-art alternatives. This benchmark provides compelling evidence for the efficacy of domain-specific finetuning. Our key contributions include:

\begin{itemize}[topsep=2pt, parsep=1ex, partopsep=1ex, itemsep=0.5ex]
\item RiskData: We introduce a specialized QA dataset designed for finetuning embedding models in risk management. Beyond finetuning, this dataset facilitates vocabulary expansion, enabling models to better capture domain-specific terminology and contextual nuances.
\item RiskEmbed: We release RiskEmbed, our finetuned embedding model, empowering researchers and practitioners to leverage its enhanced capabilities. We further provide a detailed comparative analysis, showcasing its superior accuracy against both general-purpose and financial domain-specific embedding models.
\end{itemize}

\section{Prior Work} \label{sec:pwork}
Several datasets have been developed to support general financial question-answering systems, covering various aspects of financial information retrieval and analysis. These datasets collectively enable the development of robust financial question-answering models by capturing diverse financial knowledge sources, from news reports to investment advice and structured financial data. 
However, it is important to note that these datasets are designed for general financial question-answering and do not specifically address risk management. In other words, they do not focus on the specialized needs of risk assessment, regulatory compliance, or financial risk prediction.

Trade-the-event \parencite{TradeTheEvent} and Stock News Sentiments \parencite{StockNewsSentiments} focus on corporate event news and summaries, providing insights into how events impact financial markets. FinQABench \parencite{FinQABench} and FinanceBench \parencite{FinanceBench} contain questions related to publicly traded companies and their relevant public filings, enabling models to extract and interpret financial disclosures. TAT-QA \parencite{TATQA} facilitates question-answering over hybrid tabular and textual data, reflecting the complexity of financial documents. Similarly, ConvFinQA \parencite{ConvFinQA} and FinQA \parencite{FinQA} provide question-answer pairs derived from financial reports, enhancing the system’s reasoning ability over structured financial statements. For personal and investment-related inquiries, FIQA Personal Finance \parencite{FIQA} and Finance Alpaca \parencite{FinanceAlpaca} offer datasets focusing on financial advice and personal finance Questions and answers. HC3 finance \parencite{HC3Finance} also contributes to financial advice question-answering, while News Stocks \parencite{NewsStocks} provides finance news and summaries to enrich contextual understanding. 

General financial documents, such as earnings reports and investor presentations, frequently employ distinct wordings and acronyms tailored to their respective domains such as P/E ratio (Price-to-Earnings ratio) and ROE (Return on Equity) to describe company performance and market trends. These documents are written in a way that is accessible to investors, analysts, and the general public, focusing on financial growth, profitability, and market positioning. In contrast, risk management documents, such as regulatory filings and stress testing reports, employ terminology specific to risk assessment and compliance, including VaR (Value at Risk), PD (Probability of Default), LGD (Loss Given Default), RWA (Risk-Weighted Assets), and AML (Anti-Money Laundering). These documents often contain technical language designed for regulatory bodies, auditors, and risk professionals, emphasizing risk exposure, regulatory adherence, and financial stability. The differing use of terminology and acronyms reflects the specialized focus of each document type, making them distinct in content.

\section{Dataset Generation} \label{sec:dataset}
To develop a high-quality dataset for risk management, we curated 94 guideline documents openly published by the Office of the Superintendent of Financial Institutions (OSFI) between 1991 and 2024. OSFI in Canada provides comprehensive risk management regulatory guidelines covering various aspects of financial risk, including credit risk, market risk, operational risk, and regulatory compliance. These guidelines, while tailored for Canadian banks, align closely with international frameworks and share common terms and principles with those used in other countries, including the United States. For example, OSFI’s guidelines are consistent with global standards, such as those from the Basel Committee on Banking Supervision. This consistency not only enhances regulatory cooperation but also makes it easier to develop and deploy risk management models, as the shared terminology and principles allow for a unified approach. This commonality facilitates the use of risk management models across countries, enabling more effective and accurate risk management question-answering systems on a global scale. Therefore, the dataset ensures comprehensive coverage of risk management principles and best practices, making it suitable for training and evaluating RAG models in the risk management sector.  

To construct this QA dataset tailored to risk management, we leveraged an LLM (i.e., Gemini 1 Pro) to generate 7,496 positive pairs (also known as question-context pairs).
The dataset contains structured inquiries related to financial regulations, insurance, actuarial responsibilities, structured settlements, and securitization. Each question is linked to a specific financial or regulatory topic, facilitating the finetuning of an embedding model. The dataset primarily focuses on:
\begin{itemize}
    \item Insurance Regulations and Actuarial Practices – Covers responsibilities of appointed actuaries, financial reporting implications, and structured settlements.
    \item Securitization and Risk Management – Includes questions on asset-backed securities, special purpose entities (SPEs), credit risk mitigation, and capital adequacy.
    \item Regulatory Compliance and Guidelines – Addresses OSFI and Basel framework requirements, financial institution assessments, and governance policies.
    \item Operational Risk and Financial Modeling – Discusses capital requirements, loss calculations, and operational risk categorization.
\end{itemize}

To ensure the accuracy and reliability of the dataset, all generated question-context pairs underwent manual validation by our specialized staff with expertise in risk management and regulatory compliance. Each pair was reviewed for factual correctness, alignment with OSFI guidelines, and relevance to real-world risk assessment. This validation process enhances the dataset's credibility, reducing the likelihood of hallucinated or misleading responses in downstream applications. By combining automated question generation with human verification, our dataset maintains a high standard of quality, making it a valuable resource for risk-focused retrieval and question-answering systems.  

This dataset serves as the foundation for finetuning an embedding model optimized for regulatory compliance and financial risk management. By incorporating domain-specific language and structured risk-related content, the finetuned model enhances retrieval performance within a RAG framework, improving the accuracy and relevance of responses. Open-sourcing this dataset facilitates advancements in risk-aware AI applications, enabling financial institutions to develop more robust, regulatory-aligned language models for compliance automation and risk analysis.

    
    
    


\section{Model Finetuning} \label{sec:setup}
Here, our aim is to finetune an open-source, efficient, and accurate embedding model to enhance its performance for RAG systems. Finetuning is a crucial step in adapting pre-trained models to specific tasks, allowing them to better capture the semantic nuances of the target domain. We use the Snowflake Arctic Embed-Medium model \parencite{arctic} as the base embedding model, selected for its balance of computational efficiency and retrieval accuracy. Notably, this model is among the top-performing models on the Hugging Face MTEB (Massive Text Embedding Benchmark) leaderboard \parencite{hf}, demonstrating strong retrieval capabilities across various tasks. It has 305 million parameters, an embedding dimension of 768, and achieves a retrieval NDCG@10 score of 55\% for general document retrieval, making it a competitive choice for our finetuning experiments.

For training, we construct a dataset composed of question-context pairs, where each question is associated with its most relevant textual context. To ensure robust evaluation, we split the dataset into two subsets, allocating 95\% for training and 5\% for testing. The training data is used to optimize the model’s embedding representations, while the test set serves to assess its generalization performance. Given that our dataset contains positive pairs (i.e., known relevant question-context pairs), we employ the Multiple Negatives Ranking (MNR) loss function during finetuning, as shown in Equation \ref{eq:mnr_loss} \parencite{MNR1, MNR2, MNR3}. This loss function encourages the model to generate embeddings that maximize the similarity between positive pairs while minimizing similarity with randomly sampled negative pairs. The MNR loss is particularly effective in retrieval-based tasks, as it helps the model learn to distinguish relevant contexts from irrelevant ones \parencite{whyMNR}. Note that the denominator explicitly excludes the positive example when summing over all negatives. We finetune the model using a batch size of 12 and train it for 2 epochs. Notably, experiments with additional epochs led to performance degradation, likely due to overfitting on the training data. This observation underscores the importance of carefully selecting training hyperparameters to avoid diminishing returns in finetuning efforts.

\begin{equation} \label{eq:mnr_loss}
\mathcal{L} = -\sum_{i=1}^{N} \log \frac{\exp(s(q_i, p_i))}{\exp(s(q_i, p_i)) + \sum_{j \neq i} \exp(s(q_i, n_j))}
\end{equation}
where
\begin{itemize}
    \item $N$ is the number of query-positive pairs in the batch.
    \item $q_i$ is the $i$-th query.
    \item $p_i$ is the positive document corresponding to the $i$-th query.
    \item $n_j$ is the $j$-th negative document in the batch (including all documents except $p_i$ for query $q_i$).
    \item $s(q, d)$ is the similarity score between query $q$ and document $d$.
\end{itemize}

To quantify the improvements brought by finetuning, we evaluate both the base and finetuned models using standard ranking metrics commonly used in information retrieval tasks. Specifically, we compute Mean Reciprocal Rank at 10 (MRR@10), Mean Average Precision at 100 (MAP@100), and Normalized Discounted Cumulative Gain at 10 (NDCG@10). They are key ranking metrics used in information retrieval, each emphasizing different aspects of retrieval quality. MRR@10 measures how early the first relevant document appears in the ranked list, making it particularly useful for tasks where retrieving a single correct answer quickly is crucial, such as question-answering systems. However, it does not consider multiple relevant results. NDCG@10, on the other hand, evaluates both the ranking order and relevance of retrieved documents, prioritizing highly relevant results at the top using a logarithmic discount factor. This makes it well-suited for search engines and recommendation systems, where ranking quality is critical. MAP@100 provides a broader assessment by computing precision across multiple recall levels, making it ideal for document and passage retrieval tasks that require retrieving multiple relevant results rather than just the first relevant document. While MRR@10 focuses on the earliest relevant result, NDCG@10 ensures an optimal ranking structure, and MAP@100 captures overall retrieval effectiveness across a larger set of results.

\begin{figure}[!t]
  \centering
  \includegraphics[width=1\textwidth]{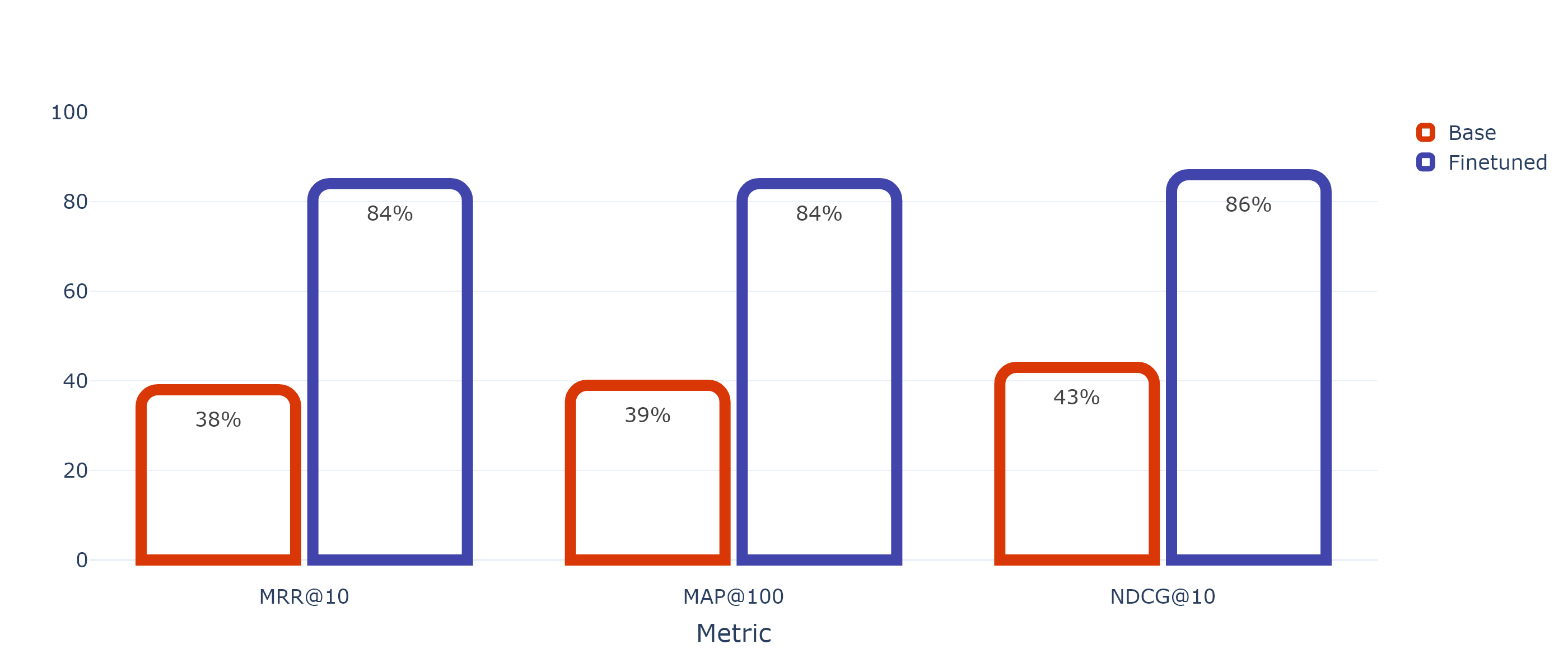}
  \caption{Performance comparison of the base and finetuned models across three retrieval metrics. The hollow bars represent the respective model performance, with the base model shown in red and the finetuned model in blue.}
  \label{fig:performance}
\end{figure}

The evaluation metrics visualized in Figure \ref{fig:performance} and summarized in Table \ref{tab:performance} demonstrate a substantial improvement in retrieval performance after finetuning. MRR@10 increased from 38\% in the base model to 84\% in the finetuned model, indicating that relevant documents are ranked much higher on average after finetuning. Similarly, NDCG@10 improved from 43\% to 86\%, showing that the finetuned model assigns significantly higher relevance scores to the most relevant retrieved documents. Additionally, MAP@100 rose from 39\% to 84\%, further reinforcing the effectiveness of finetuning in improving ranking quality over a broader set of retrieved results. The consistent and significant gains across all three metrics highlight the effectiveness of the finetuning process, demonstrating that the adapted model provides more accurate and contextually relevant embeddings compared to the base model.

\begin{table}[!t]
\centering
\caption{Performance comparison of the base and finetuned models across three retrieval metrics.}
\label{tab:performance}
\begin{tabular}{lcc}
\toprule
 & Base (Snowflake Arctic Embed-Medium) & Finetuned \\
\midrule
MRR@10 & 38\% & 84\% \\
MAP@100 & 39\% & 84\% \\
NDCG@10 & 43\% & 86\% \\
\bottomrule
\end{tabular}
\end{table}

\section{Benchmarking and Discussion} \label{sec:results}
To assess the performance of our finetuned embedding model, we benchmark it against several high-performance embedding models available via API, as these models are predominantly closed-source. This comparison provides a broader context for evaluating the effectiveness of our approach relative to state-of-the-art alternatives. Additionally, we include a finetuned model trained on general financial data to investigate whether embeddings derived from finance in general differ from those specialized in a subfield like risk management, which is the focus of this study. This allows us to examine the impact of domain specificity on embedding quality and downstream task performance. All models are evaluated on our prepared RiskData dataset. It is important to note that our finetuned model has not been exposed to this test data during training, ensuring that the evaluation reflects the model’s ability to generalize to unseen risk-related financial texts. The comparative analysis provides valuable insights into the suitability of various embedding models for risk management applications and the benefits of finetuning in specialized financial domains.

\begin{figure}[!t]
  \centering
  \includegraphics[width=1\textwidth]{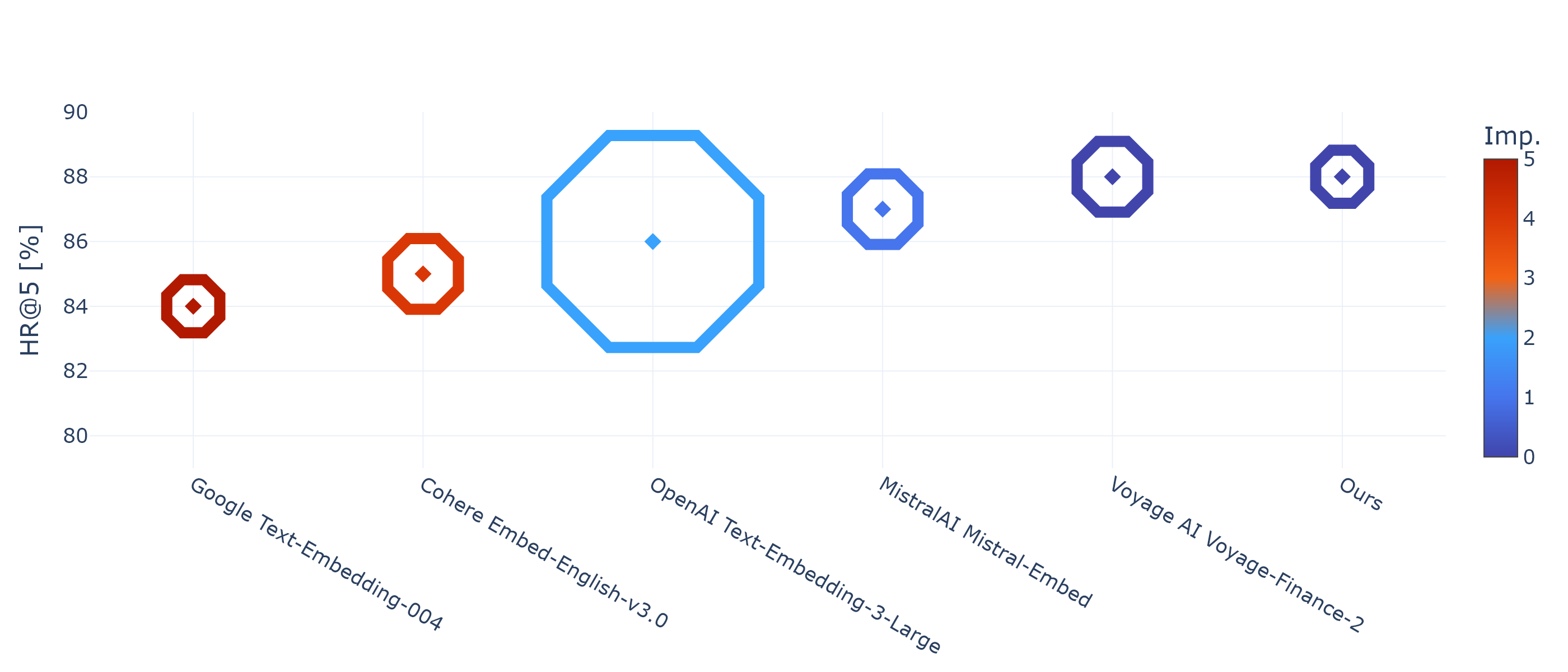}
  \caption{Benchmark visualization: It compares text embedding models, where the size of each circle corresponds to the model's embedding size. The circle colors represent our model's improvement compared to benchmark models, with redder shades signifying higher improvement and bluer shades indicating lower or no improvement.}
  \label{fig:benchmark}
\end{figure}

The benchmark results, outlined in Table \ref{tab:benchmark}, include a key performance metric Hit Rate at 5 (HR@5), our model's improvement compared to benchmark models, and embedding vector size for each model; the corresponding visualization is presented in Figure \ref{fig:benchmark}. HR@5 measures the percentage of times the correct result appears in the top five, indicating a model's ranking effectiveness, with higher values signifying better retrieval quality. This metric is intuitive and particularly useful in practical scenarios, where the focus is on top results. Compared to other metrics such as MRR, MAP, and NDCG, HR is simpler and more focused on ensuring at least one highly relevant result in the top five, making it ideal for closed-source API models. In addition, the improvement column highlights the relative gains of our finetuned model compared to the closed-source model, demonstrating the advantage of domain-specific finetuning. Furthermore, the table provides information on the varying embedding sizes across models; the embedding vector size reflects a model’s capacity to encode semantic information and larger vectors generally capture richer features but require more computational resources.

\begin{table}[!t]
\caption{Benchmark results: It details the performance of our finetuned model compared to other closed embedding models. The table shows HR@5, improvement (our model's improvement compared to benchmark models), and embedding size, illustrating our model's efficiency and effectiveness.}
\centering
\begin{tabular}{llll}
\toprule
 & HR@5 [\%] & Improvement [\%] & Embedding Size \\
\midrule
Google Text-Embedding-004 & 84 & 5 & 768 \\
Cohere Embed-English-v3.0 & 85 & 4 & 1024 \\
OpenAI Text-Embedding-3-Large & 86 & 2 & 3072 \\
MistralAI Mistral-Embed & 87 & 1 & 1024 \\
VoyageAI Voyage-Finance-2 & 88 & 0 & 1024 \\
Ours & 88 & - & 768 \\
\bottomrule
\end{tabular}
\label{tab:benchmark}
\end{table}

Our finetuned embedding model achieves state-of-the-art performance (88\%) among closed-source models. In particular, our model outperforms Google Text-Embedding-004 (84\%), Cohere Embed-English-v3.0 (85\%), OpenAI Text-Embedding-3-Large (86\%), and MistralAI Mistral-Embed (87\%), all of which were not finetuned in domain-specific data. This result highlights the advantage of finetuning on risk management data, as our model surpasses general-purpose embeddings in retrieval effectiveness. Furthermore, despite having the smallest embedding size (768 dimensions, equal to Google’s model but significantly smaller than OpenAI’s 3072 dimensions), our model efficiently encodes domain-specific information without requiring a larger vector space. Compared to VoyageAI's Voyage-Finance-2, which is also finetuned but on general financial data, our model achieves the same HR@5 (88\%). The ability to achieve peak performance with a more compact representation (768 vs. 1024 dimensions for VoyageAI) suggests that our model captures risk-related semantics more effectively. These findings reinforce the importance of finetuning for specialized tasks and demonstrate that a targeted adaptation strategy can yield optimal results without increasing computational overhead. A key advantage of our approach lies in its relatively compact embedding size (768 dimensions), which is smaller than the Voyage-Finance-2 model (1024 dimensions). This reduction in dimensionality not only decreases memory requirements but also speeds up inference — an especially critical factor for real-time applications such as RAG. Consequently, our domain-specific finetuning on risk management data demonstrates that smaller, specialized models can rival or surpass the performance of larger, more general embeddings while maintaining computational efficiency and facilitating faster deployment.

\section{Conclusion}
In this paper, we introduced RiskData, a specialized dataset for risk management question-answering, and RiskEmbed, a finetuned embedding model optimized for financial risk-related retrieval tasks. By leveraging a RAG framework, we demonstrated that finetuning on domain-specific regulatory data significantly improves retrieval accuracy compared to general-purpose embedding models. Benchmarking results against leading closed-source and financial-specific models confirmed that RiskEmbed achieves state-of-the-art performance while maintaining computational efficiency. The open-sourcing of RiskData and RiskEmbed aims to foster further advancements in AI-driven risk management, enabling more precise and efficient information retrieval for financial institutions.
For future work, we plan to explore using triplet loss and negative mining, which will enhance model robustness by improving the distinction between relevant and irrelevant contexts \parencite{nm}.
Additionally, we aim to update the tokenizer's vocabulary with risk-specific terminology and conduct further finetuning to improve retrieval accuracy \parencite{exbert}.
Expanding RiskData to include additional regulatory sources, such as international banking guidelines, could also be a focus to evaluate the generalization of the model in different financial systems and compliance frameworks.

\section*{Acknowledgments}
The authors would like to acknowledge the valuable contributions of the Risk Management team at TD Bank for their expertise in regulatory frameworks, financial risk assessment, and compliance practices, which were instrumental in the development of RiskData and the finetuning of RiskEmbed.


\printbibliography
\end{document}